\documentclass{article}
\usepackage{spconf,amsmath,graphicx,hyperref}
\usepackage{booktabs,multirow,array}
\title{Tracing and Relearning Detection Evidence in Text-to-Speech Systems}
\name{Eunji Shin$^1$, Kyudan Jung$^2$, Jihwan Kim$^2$, Minwoo Lee$^2$, Jaegul Choo$^2$}
\address{$^1$Ewha W. University, $^2$KAIST AI}
\begin{document}
\maketitle
\begin{abstract}
Recent audio deepfake detectors separate bona fide speech from synthetic speech, yet it remains unclear which stage of a text-to-speech system supplies the detection evidence. We address this with controlled resynthesis and detector adaptation in an F5-TTS--BigVGAN pipeline. Since vocoder reconstruction of a real mel can itself be separable from the source utterance, we fix the vocoder and trace the larger change in detector separation to acoustic generation. Adversarially fine-tuning the acoustic model, with no detector in its objective, raises EER against fixed detectors at comparable quality. However, adapting a detector only on the tuned model's VCTK outputs lowers its LibriSpeech EER from 19.42\% to 7.46\% and improves detection of unseen base F5-TTS outputs. These results suggest that acoustic-model updates can reduce the detection evidence available to fixed detectors, while detector adaptation keeps the updated outputs detectable in this pipeline.
\end{abstract}
\section{Introduction}
\begin{figure}[t]
\centering
\includegraphics[width=\columnwidth]{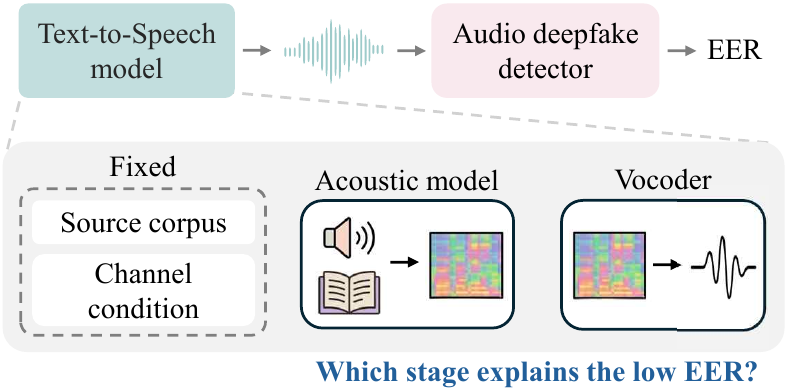}
\caption{Isolating which TTS stage supplies the detection evidence, with the source corpus and channel condition fixed.}
\label{fig:task}
\vspace{-5pt}
\end{figure}

Audio deepfake detectors report low equal error rates (EER) on standard benchmarks. These benchmarks vary the acoustic model and waveform generation across spoofing systems, and the source corpus and channel condition across benchmarks and trials~\cite{todisco2019asvspoof,yamagishi2021asvspoof}. A low EER shows that a detector separated synthetic from bona fide speech under a protocol, but not which of these factors it used. Detectors can exploit silence-duration cues unrelated to synthesis~\cite{muller2021silence}, and vocoder reconstruction of a real mel spectrogram can itself be separable from the source utterance~\cite{sun2023ai}. In a TTS output the vocoder receives a mel generated by the acoustic model rather than extracted from speech, so a benchmark EER does not attribute the separation to either stage, motivating controls that isolate which stage of the pipeline a low EER reflects (Fig.~\ref{fig:task}).

Modern text-to-speech (TTS) systems split synthesis into two stages, an acoustic model that generates mel spectrograms and a vocoder that converts them into waveforms. In F5-TTS~\cite{chen2025f5}, the acoustic model is a flow-matching Diffusion Transformer (DiT)~\cite{peebles2023scalable}, and its checkpoint supports BigVGAN~\cite{lee2023bigvgan} and Vocos~\cite{siuzdak2024vocos} as its vocoder. Passing real and generated mels through the same vocoder removes it as a difference between a resynthesized real mel and a generated one. We fix BigVGAN, which leaves the least separation at reconstruction of the two (Table~\ref{tab:vocoder}) and whose reconstructions serve as the real target in Sec.~\ref{sec:finetuning}. The larger part of the remaining separation is then attributable to acoustic generation. We update the acoustic model alone, fine-tuning it without any evaluation detector in its objective. These outputs raise EER against fixed detectors, so acoustic generation is a stage where the detection evidence can be weakened.

However, adapting a detector to the updated outputs improves detection of both these outputs and those from the base model, even though the latter were not used for adaptation. In this pipeline, a higher EER against fixed detectors after a generator update therefore need not mean that the generated speech has become indistinguishable from bona fide speech.

Our contributions are as follows:
\vspace{-5pt}
\begin{itemize}
\setlength{\itemsep}{1pt}
\setlength{\parskip}{0pt}
\item We use a fixed-vocoder resynthesis comparison to trace the detection evidence to acoustic generation under BigVGAN and suggest that the share left for this stage depends on the vocoder (Tables~\ref{tab:main_eer}--\ref{tab:vocoder}).
\item We show that acoustic-model fine-tuning increases EER across fixed detectors on LibriSpeech and VCTK beyond RMS gain matching, while quality and content measures stay comparable (Tables~\ref{tab:main_eer}--\ref{tab:quality}).
\item We show that detector adaptation learns the tuned outputs and transfers to base F5-TTS, while legacy EER differs between adaptation corpora (Table~\ref{tab:detector_adapt}).
\end{itemize}

\begin{figure*}[t]
\centering
\includegraphics[width=0.95\textwidth]{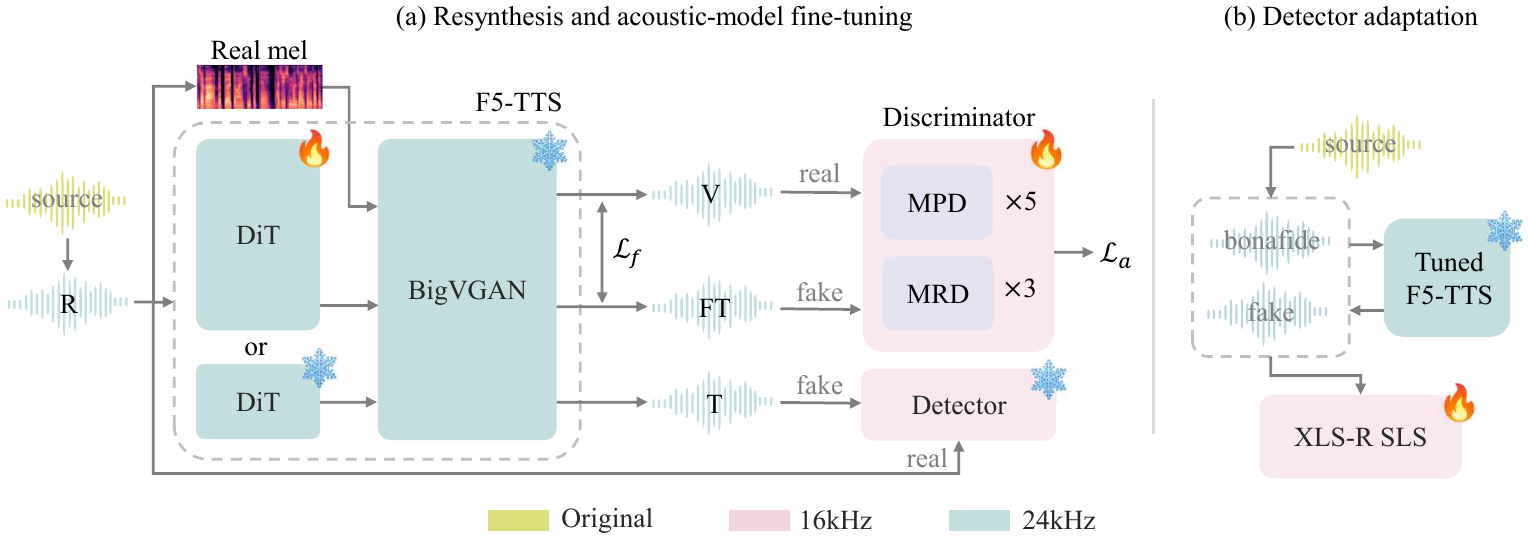}
\caption{Overview of the experimental framework, with conditions defined in Sec.~\ref{sec:conditions}. (a) $\mathcal{L}_a$ and $\mathcal{L}_f$ denote the adversarial and feature matching losses. When R itself is evaluated, the original source serves as bona fide. (b) XLS-R SLS is adapted on bona fide utterances paired with their tuned F5-TTS resyntheses.}
\label{fig:architecture}
\end{figure*}

\vspace{-10pt}
\section{Method}
\vspace{-5pt}
\subsection{Paired resynthesis and controls}
\label{sec:conditions}
\vspace{-5pt}
Figure~\ref{fig:architecture}(a) outlines the resynthesis pipeline and our fine-tuning. We define five conditions for the same source utterance and target transcript, prompting every generated evaluation condition with a held-out utterance of the same speaker rather than the one being resynthesized. \textbf{R} resamples the source to 24\,kHz without synthesis. \textbf{V} extracts a deterministic real mel from the source utterance and reconstructs it with BigVGAN, the vocoded spoof construction of~\cite{wang2023spoofed} used here for evaluation rather than training. \textbf{T} generates a mel spectrogram with the base F5-TTS DiT and uses the same BigVGAN. \textbf{FT} replaces the base DiT with one fine-tuned as in Sec.~\ref{sec:finetuning}, and is otherwise identical to T; the vocoder, text processing, prompts, and sampling steps remain fixed. \textbf{TG} applies a scalar gain to each T output to match the RMS amplitude of its paired FT output, a level the fine-tuning objective does not constrain.

The R--V difference is attributable to BigVGAN reconstruction. V--T keeps the same vocoder and replaces the source mel with a mel the DiT generates from text. Because the source mel also carries the prosody and acoustic realization of the source utterance, V--T covers these differences as well as any detection evidence from generation. T--TG isolates the share of the T--FT gap that level alone can produce.

\subsection{Acoustic-model fine-tuning}
\label{sec:finetuning}
We fine-tune all 337.1 million DiT parameters, freezing BigVGAN and the remaining TTS components. BigVGAN stays differentiable, so the discriminator loss computed on the waveform reaches the DiT through it. The jointly trained discriminator starts from random initialization and combines five multi-period discriminators (periods 2, 3, 5, 7, 11) with three multi-resolution discriminators (Fourier sizes 2048, 1024, 512)~\cite{kumar2023high}. Its real targets are BigVGAN reconstructions of real mels; its fake inputs are BigVGAN outputs from DiT-generated mels. Sharing the vocoder removes vocoder identity as a difference between discriminator inputs, so the only difference is where the mel came from. Following the least-squares adversarial and feature-matching formulation of HiFi-GAN~\cite{kong2020hifigan}, we weight these losses by one and two, respectively. We use no mel reconstruction loss, and feature matching uses absolute activation differences. The evaluation detectors remain outside this objective.

\subsection{Detector adaptation}
A higher EER on a fixed detector leaves two readings open: the tuned output may carry no detection evidence left to catch, or it may carry evidence unseen by the detector. To tell them apart, we start from the released XLS-R SLS~\cite{zhang2024sls} checkpoint and keep training it with FT labeled as synthetic (Fig.~\ref{fig:architecture}(b)). In the adaptation set only, each FT example is paired with the bona fide utterance that supplied its prompt and transcript, so content and speaker cannot stand in for the label. Neither adaptation run includes T examples, so transfer to base F5-TTS is measured rather than trained for. Evaluation covers the held-out conditions of Sec.~\ref{sec:conditions} and legacy ASVspoof 2021 DF/LA~\cite{yamagishi2021asvspoof} and In-the-Wild~\cite{muller2022generalize} benchmarks.

\section{Experimental Setup}
\subsection{Datasets}
LibriSpeech~\cite{panayotov2015librispeech} is used for acoustic-model fine-tuning, detector adaptation, and evaluation. Acoustic-model fine-tuning uses train-clean-100. The LibriSpeech adaptation run draws 2580 utterances from 20 speakers of that split; its bona fide speech is resampled via 24\,kHz and receives a constant level correction, so loudness cannot stand in for the label. Evaluation uses 800 test-clean utterances from 40 other speakers.

VCTK~\cite{yamagishi2019vctk} is used for detector adaptation and evaluation. The VCTK adaptation run uses 2578 bona fide utterances from 20 speakers of the ASVspoof 2019~\cite{todisco2019asvspoof} training partition, and evaluation uses 1354 segments from 68 speakers that appear in neither that partition nor the ASVspoof development set, where a segment is two utterances joined to cover the detector window. Each synthetic condition generates the two halves separately and joins them, so the join cannot act as a class cue. Its bona fide and FT levels differ by 0.09\,dB, so this run needs no level correction. VCTK lies within the bona fide domain the detectors were trained on, since ASVspoof 2019 draws its bona fide speech from it. An additional 800 English VoxPopuli utterances of parliamentary speech from 40 speakers~\cite{wang2021voxpopuli} are used for evaluation only.
\vspace{-5pt}

\subsection{Training, detectors, and metrics}
The acoustic-model update runs at DiT and discriminator learning rates of 10$^{-6}$ and 10$^{-5}$ with 200 warmup updates and a batch size of eight across four NVIDIA A100 GPUs (40\,GB each); the sampler uses 32 unrolled Euler steps. The reported checkpoint is at 1800 updates. It is the last of ten saved at 200-update intervals whose UTMOS, speaker similarity, and WER on LibriSpeech test-clean are no worse than the base model's. Detector EER is not part of this criterion. Both adaptation runs train for 2000 updates with a batch size of 14 on one NVIDIA A100 GPU (40\,GB) and three training seeds, and each adapted detector is evaluated on both corpora.

We use three fixed detectors trained on ASVspoof 2019 logical access data. XLS-R SLS (SLS)~\cite{zhang2024sls} and XLSR-Mamba (Mamba)~\cite{xiao2025xlsr} share an XLS-R 300M~\cite{babu2022xlsr} front end and differ in their back ends, while AASIST-L~\cite{jung2022aasist} reads the waveform through a sinc-convolution front end. EER is reported in percent, and movement toward 50\% indicates poorer separation under the fixed score orientation.

We evaluate UTMOS~\cite{baba2024t05}, ECAPA-TDNN cosine speaker similarity to the prompt utterance~\cite{desplanques2020ecapa}, and Whisper large-v3 word error rate (WER)~\cite{radford2023robust}, micro-averaged over reference words. T, TG, and FT use seeds (1234, 42, 913); means and standard deviations summarize three corpus-level scores, while R and V are deterministic.

\section{Results and Analysis}
\subsection{Separation emerges after acoustic generation under BigVGAN}
\begin{table}[t]
\centering
\caption{EER of fixed detectors per condition. For V, T, TG, and FT, the bona fide trials are 24 kHz resampled sources.}
\label{tab:main_eer}
\resizebox{0.91\columnwidth}{!}{%
\setlength{\tabcolsep}{3pt}
\begin{tabular}{@{}llccc@{}}
\toprule
\textbf{Corpus} & \textbf{Cond.} & \textbf{SLS} & \textbf{Mamba} & \textbf{AASIST-L} \\
\midrule
\multirow{5}{*}{LibriSpeech} & R & 45.25 & 47.88 & 49.88 \\
& V & 38.00 & 38.63 & 50.63 \\
& T & 15.46$_{\pm 0.44}$ & 12.38$_{\pm 0.78}$ & 28.75$_{\pm 1.11}$ \\
& TG & 17.50$_{\pm 0.45}$ & 13.46$_{\pm 0.69}$ & 40.88$_{\pm 0.99}$ \\
& FT & 19.42$_{\pm 1.39}$ & 16.04$_{\pm 0.81}$ & 47.67$_{\pm 1.18}$ \\
\midrule
\multirow{5}{*}{VCTK} & R & 51.26 & 49.93 & 49.93 \\
& V & 45.94 & 45.27 & 51.33 \\
& T & 3.45$_{\pm 0.19}$ & 4.36$_{\pm 0.07}$ & 28.95$_{\pm 0.84}$ \\
& TG & 3.32$_{\pm 0.13}$ & 4.36$_{\pm 0.15}$ & 29.52$_{\pm 0.17}$ \\
& FT & 4.09$_{\pm 0.17}$ & 5.96$_{\pm 0.11}$ & 31.71$_{\pm 0.83}$ \\
\bottomrule
\end{tabular}%
}
\vspace{-10pt}
\end{table}
\begin{table}[t]
\centering
\caption{SLS EER for real-mel reconstruction (V) with the two vocoder configurations the F5-TTS checkpoint supports.}
\vspace{+1pt}
\label{tab:vocoder}
\resizebox{0.58\columnwidth}{!}{%
\begin{tabular}{@{}lcc@{}}
\toprule
\textbf{Vocoder} & \textbf{LibriSpeech} & \textbf{VCTK} \\
\midrule
BigVGAN & 38.00 & 45.94 \\
Vocos & 12.00 & 22.01 \\
\bottomrule
\end{tabular}%
}
\vspace{-10pt}
\end{table}

Table~\ref{tab:main_eer} reports EER for every condition. R stays near chance on both corpora, so resampling the source produces little separation by itself. V lowers EER for SLS and Mamba, and the further decrease from V to T is roughly 3 times the R--V step on LibriSpeech and 8 to 9 times on VCTK.

AASIST-L reads the waveform through a different front end, so the larger V–T than R–V contrast does not rest on the XLS-R features that SLS and Mamba share. For this detector both R and V sit at chance on either corpus, and its whole decrease appears at T. 

The contrast therefore holds across front ends and back ends, while the amount of separation depends on the detector and the corpus. With BigVGAN, the separation appears mainly after the DiT generates the mel. The V--T comparison does not distinguish the prosody and acoustic realization of the generated utterance from any artifact of generation.

For SLS and Mamba, V reproduces the separation reported for reconstruction alone~\cite{sun2023ai}, and its size is vocoder-dependent: for SLS, Vocos reconstructions are separated far more strongly than BigVGAN's on both corpora (Table~\ref{tab:vocoder}).

\subsection{Fine-tuning raises EER beyond RMS gain matching}
FT increases EER over T in all six detector--corpus pairs in Table~\ref{tab:main_eer}. The increase exceeds the spread across generation seeds in every pair, ranging from 0.64 percentage points for SLS on VCTK to 18.92 points for AASIST-L on LibriSpeech, where it leaves the detector near chance. The direction holds for all three detectors, so the acoustic-model update changes evidence that all of them depend on. At the same time, the low FT EERs on VCTK for SLS and Mamba show that the tuned outputs remain detectable.

TG isolates the part of the T--FT increase that output level alone can produce. On LibriSpeech, RMS gain matching accounts for 52\% of the T--FT gap for SLS and 30\% for Mamba. On VCTK, TG leaves SLS and Mamba nearly unchanged, whereas FT increases both. AASIST-L is more level-sensitive: on LibriSpeech, TG accounts for 64\% of the gap. A residual above TG remains in all six pairs, although the comparison does not say whether local amplitude, spectral, or temporal properties account for it.

\begin{table}[t]
\centering
\caption{SLS EER alongside quality and content measures for T and FT. Higher UTMOS and speaker similarity and lower WER (\%) are better.}
\label{tab:quality}
\resizebox{\columnwidth}{!}{%
\begin{tabular}{@{}llcccc@{}}
\toprule
\textbf{Corpus} & \textbf{Cond.} & \textbf{SLS EER} & \textbf{UTMOS} & \textbf{Spk. sim.} & \textbf{WER} \\
\midrule
\multirow{2}{*}{LibriSpeech} & T & 15.46$_{\pm 0.44}$ & 3.38$_{\pm 0.02}$ & 0.724$_{\pm 0.002}$ & 2.98$_{\pm 0.12}$ \\
& FT & 19.42$_{\pm 1.39}$ & 3.41$_{\pm 0.01}$ & 0.725$_{\pm 0.001}$ & 2.95$_{\pm 0.03}$ \\
\midrule
\multirow{2}{*}{VCTK} & T & 3.45$_{\pm 0.19}$ & 3.37$_{\pm 0.03}$ & 0.754$_{\pm 0.002}$ & 0.73$_{\pm 0.02}$ \\
& FT & 4.09$_{\pm 0.17}$ & 3.35$_{\pm 0.03}$ & 0.763$_{\pm 0.004}$ & 0.68$_{\pm 0.02}$ \\
\midrule
\multirow{2}{*}{VoxPopuli} & T & 35.33$_{\pm 1.49}$ & 3.35$_{\pm 0.04}$ & 0.747$_{\pm 0.000}$ & 1.41$_{\pm 0.07}$ \\
& FT & 40.75$_{\pm 0.99}$ & 3.45$_{\pm 0.03}$ & 0.742$_{\pm 0.004}$ & 1.46$_{\pm 0.08}$ \\
\bottomrule
\end{tabular}%
}
\vspace{-10pt}
\end{table}

Table~\ref{tab:quality} places the EER changes alongside quality and content measures. Across the three corpora, the changes are small in both directions: each measure improves on some corpora and worsens on others, and the largest degradation on any measure is within or close to the spread across generation seeds. The EER increase therefore does not come from degrading the output, on two corpora that played no part in selecting the checkpoint. These measures leave open whether acoustic properties they do not capture changed.

\subsection{Adaptation learns tuned outputs}
Table~\ref{tab:detector_adapt} reports EER after each adaptation run. LibriSpeech adaptation lowers FT EER from 19.42\% to 1.71\% on LibriSpeech and from 4.09\% to 0.20\% on VCTK, and VCTK adaptation lowers it to 7.46\% on LibriSpeech and 0.10\% on VCTK. T falls to comparable EER in both runs, though adaptation labels only FT as synthetic. Either corpus therefore teaches the detector to separate FT, and the FT outputs that raised EER against the fixed detectors are the ones separated at these rates. The higher EER of FT against a fixed detector therefore does not mean that nothing was left to catch.

\begin{table}[t]
\centering
\caption{SLS adaptation: EER (lower is better for T/TG/FT). LS/VC: balanced LibriSpeech and VCTK adaptation (3 training seeds). Baseline T/TG/FT summarize generation seeds. $^*$Released checkpoint~\cite{zhang2024sls}}
\label{tab:detector_adapt}
\resizebox{0.48\textwidth}{!}{%
\begin{tabular}{@{}llccc@{}}
\toprule
\textbf{Eval. set} & \textbf{Cond.} & \textbf{Baseline} & \textbf{LS adapt.} & \textbf{VC adapt.} \\
\midrule
\multirow{5}{*}{LibriSpeech} & R & 45.25 & 49.50$_{\pm 0.38}$ & 46.50$_{\pm 0.33}$ \\
& V & 38.00 & 40.79$_{\pm 0.64}$ & 35.46$_{\pm 0.88}$ \\
& T & 15.46$_{\pm 0.44}$ & 1.42$_{\pm 0.44}$ & 5.92$_{\pm 0.47}$ \\
& TG & 17.50$_{\pm 0.45}$ & 1.75$_{\pm 0.57}$ & 6.29$_{\pm 0.07}$ \\
& FT & 19.42$_{\pm 1.39}$ & 1.71$_{\pm 0.26}$ & 7.46$_{\pm 0.31}$ \\
\midrule
\multirow{5}{*}{VCTK} & R & 51.26 & 50.02$_{\pm 0.04}$ & 50.12$_{\pm 0.21}$ \\
& V & 45.94 & 46.65$_{\pm 0.43}$ & 36.34$_{\pm 1.79}$ \\
& T & 3.45$_{\pm 0.19}$ & 0.15$_{\pm 0.07}$ & 0.07$_{\pm 0.00}$ \\
& TG & 3.32$_{\pm 0.13}$ & 0.22$_{\pm 0.15}$ & 0.07$_{\pm 0.00}$ \\
& FT & 4.09$_{\pm 0.17}$ & 0.20$_{\pm 0.04}$ & 0.10$_{\pm 0.04}$ \\
\midrule
ASV21 DF & -- & 1.92$^*$ & 3.20$_{\pm 0.17}$ & 1.90$_{\pm 0.10}$ \\
ASV21 LA & -- & 2.87$^*$ & 8.35$_{\pm 0.19}$ & 3.35$_{\pm 0.18}$ \\
In-the-Wild & -- & 7.46$^*$ & 10.66$_{\pm 0.87}$ & 9.53$_{\pm 0.19}$ \\
\bottomrule
\end{tabular}%
}
\vspace{-5pt}
\end{table}
R and V test whether the adapted detector separates resampling or vocoding alone. After either adaptation run, R stays near chance and V stays far above T and FT. The detector therefore does not learn to separate resampled audio, but VCTK adaptation lowers V EER on both corpora, so it partly separates BigVGAN output from a real mel.

The two adaptation runs differ in how much of the detector's performance on legacy benchmarks they retain. VCTK adaptation stays close to the released checkpoint on ASVspoof 2021 DF but rises on LA and In-the-Wild. LibriSpeech adaptation is above the VCTK run on all three benchmarks. The run that reaches the lower EER on LibriSpeech T and FT therefore retains less legacy performance. Because both runs use adaptation sets of nearly equal size and the same number of updates, we attribute this difference to the adaptation corpus. Of the two corpora, only VCTK lies within the bona fide domain of the detector's training data. However, switching the corpus changes both the bona fide utterances and the synthetic outputs paired with them, so this experiment cannot distinguish the effect of the bona fide domain from that of the synthetic outputs.

\noindent\textbf{Joint interpretation.}
Tables~\ref{tab:main_eer} and~\ref{tab:detector_adapt} record two changes in opposite directions. On LibriSpeech, fine-tuning the DiT raises SLS EER on its own outputs, and adapting the detector on VCTK then lowers that EER to well below where the base outputs started. Read together with the larger V--T than R--V contrast, these changes suggest that the detection evidence appearing at acoustic generation changes with an acoustic-model update, and that a detector can still learn to distinguish the updated outputs through adaptation.

\noindent\textbf{Scope.} This study covers one acoustic model with one vocoder and adapts one detector architecture. The reported standard deviations reflect generation and adaptation seeds rather than speaker variation.

\section{Conclusion}
In the F5-TTS--BigVGAN pipeline, our controlled comparisons show that acoustic generation produces a larger change in detector separation than real-mel reconstruction. Fine-tuning the acoustic model against a discriminator that treats real-mel reconstructions as real makes its outputs harder for fixed detectors to identify, while quality and content measures stay comparable. To determine whether these outputs remain distinguishable, we adapt XLS-R SLS to them and find improved detection of both fine-tuned outputs and base-model outputs unseen during adaptation. This recovery shows that the reduced effectiveness of fixed detectors does not imply that detection evidence has disappeared, supporting adaptation as a way to keep pace with acoustic-model updates.

\bibliographystyle{IEEEbib}
\bibliography{refs}
\end{document}